\documentclass[twocolumn]{aastex63}

\shorttitle{{\it TESS} Observations of Known Exoplanet Hosts}
\shortauthors{Stephen R. Kane et al.}

\begin{document}

\title{Science Extraction from {\it TESS} Observations of Known
  Exoplanet Hosts}


\author[0000-0002-7084-0529]{Stephen R. Kane}
\affiliation{Department of Earth and Planetary Sciences, University of
  California, Riverside, CA 92521, USA}
\email{skane@ucr.edu}

\author[0000-0003-4733-6532]{Jacob L. Bean}
\affil{Department of Astronomy \& Astrophysics, University of Chicago,
  5640 South Ellis Avenue, Chicago, IL 60637, USA}

\author[0000-0002-4588-5389]{Tiago L. Campante}
\affiliation{Instituto de Astrof\'{\i}sica e Ci\^{e}ncias do
  Espa\c{c}o, Universidade do Porto, Rua das Estrelas, 4150-762 Porto,
  Portugal}
\affiliation{Departamento de F\'{\i}sica e Astronomia, Faculdade de
  Ci\^{e}ncias da Universidade do Porto, Rua do Campo Alegre, s/n,
  4169-007 Porto, Portugal}

\author[0000-0002-4297-5506]{Paul A. Dalba}
\altaffiliation{NSF Astronomy and Astrophysics Postdoctoral Fellow}
\affiliation{Department of Earth and Planetary Sciences, University of
  California, Riverside, CA 92521, USA}

\author[0000-0002-3551-279X]{Tara Fetherolf}
\affiliation{Department of Earth and Planetary Sciences, University of
  California, Riverside, CA 92521, USA}

\author[0000-0003-4603-556X]{Teo Mocnik}
\affiliation{Gemini Observatory, Northern Operations Center, 670
  N. A'ohoku Place, Hilo, HI 96720, USA}

\author[0000-0001-7968-0309]{Colby Ostberg}
\affiliation{Department of Earth and Planetary Sciences, University of
  California, Riverside, CA 92521, USA}

\author[0000-0002-3827-8417]{Joshua Pepper}
\affiliation{Department of Physics, Lehigh University, 16 Memorial
  Drive East, Bethlehem, PA 18015, USA}

\author[0000-0003-0447-9867]{Emilie R. Simpson}
\affiliation{Department of Earth and Planetary Sciences, University of
  California, Riverside, CA 92521, USA}

\author[0000-0002-0569-1643]{Margaret C. Turnbull}
\affiliation{SETI Institute, Carl Sagan Center for the Study of Life
  in the Universe, Off-Site: 2613 Waunona Way, Madison, WI 53713, USA}


\author[0000-0003-2058-6662]{George R. Ricker}
\affiliation{Department of Physics and Kavli Institute for
  Astrophysics and Space Research, Massachusetts Institute of
  Technology, Cambridge, MA 02139, USA}

\author[0000-0001-6763-6562]{Roland Vanderspek}
\affiliation{Department of Physics and Kavli Institute for
  Astrophysics and Space Research, Massachusetts Institute of
  Technology, Cambridge, MA 02139, USA}

\author[0000-0001-9911-7388]{David W. Latham}
\affiliation{Center for Astrophysics ${\rm \mid}$ Harvard {\rm \&}
  Smithsonian, 60 Garden Street, Cambridge, MA 02138, USA}

\author[0000-0002-6892-6948]{Sara Seager}
\affiliation{Department of Physics and Kavli Institute for
  Astrophysics and Space Research, Massachusetts Institute of
  Technology, Cambridge, MA 02139, USA}
\affiliation{Department of Earth, Atmospheric and Planetary Sciences,
  Massachusetts Institute of Technology, Cambridge, MA 02139, USA}
\affiliation{Department of Aeronautics and Astronautics, MIT, 77
  Massachusetts Avenue, Cambridge, MA 02139, USA}

\author[0000-0002-4265-047X]{Joshua N. Winn}
\affiliation{Department of Astrophysical Sciences, Princeton
  University, 4 Ivy Lane, Princeton, NJ 08544, USA}

\author[0000-0002-4715-9460]{Jon M. Jenkins}
\affiliation{NASA Ames Research Center, Moffett Field, CA, 94035, USA}


\author[0000-0001-8832-4488]{Daniel Huber}
\affiliation{Institute for Astronomy, University of Hawai`i, 2680
  Woodlawn Drive, Honolulu, HI 96822, USA}

\author[0000-0002-5714-8618]{William J. Chaplin}
\affiliation{Stellar Astrophysics Centre (SAC), Department of Physics
  and Astronomy, Aarhus University, Ny Munkegade 120, DK-8000 Aarhus
  C, Denmark}
\affiliation{School of Physics and Astronomy, University of
  Birmingham, Birmingham B15 2TT, UK}


\begin{abstract}

The transit method of exoplanet discovery and characterization has
enabled numerous breakthroughs in exoplanetary science. These include
measurements of planetary radii, mass-radius relationships, stellar
obliquities, bulk density constraints on interior models, and
transmission spectroscopy as a means to study planetary
atmospheres. The Transiting Exoplanet Survey Satellite ({\it TESS})
has added to the exoplanet inventory by observing a significant
fraction of the celestial sphere, including many stars already known
to host exoplanets. Here we describe the science extraction from {\it
  TESS} observations of known exoplanet hosts during the primary
mission. These include transit detection of known exoplanets,
discovery of additional exoplanets, detection of phase signatures and
secondary eclipses, transit ephemeris refinement, and asteroseismology
as a means to improve stellar and planetary parameters. We provide the
statistics of {\it TESS} known host observations during Cycle 1 \& 2,
and present several examples of {\it TESS} photometry for known host
stars observed with a long baseline. We outline the major discoveries
from observations of known hosts during the primary mission. Finally,
we describe the case for further observations of known exoplanet hosts
during the {\it TESS} extended mission and the expected science yield.

\end{abstract}



\section{Introduction}
\label{intro}

Discoveries of exoplanets have increased dramatically over the past
two decades, largely due to the implementation of the transit method
\citep{borucki1984a,hubbard2001}. In particular, space-based
photometry combined with large-scale survey strategies are able to
overcome both the transit probability distribution and the
observational window function that can impede ground-based approaches
\citep{kane2008b,vonbraun2009}. Significant contributors to the
space-based transit survey approaches have been the Convection,
Rotation and planetary Transits ({\it CoRoT}) mission
\citep{auvergne2009} and the {\it Kepler} mission
\citep{borucki2010a}. In 2018, the Transiting Exoplanet Survey
Satellite ({\it TESS}) was launched to begin its transit survey of the
nearest and brightest stars \citep{ricker2015}. The advantage of such
bright stars is their suitability for follow-up observations that
measure planetary masses \citep{fischer2016,burt2018} and atmospheric
compositions via transmission spectroscopy
\citep{seager2000b,kempton2018}. Each of the first two years of the
      {\it TESS} mission were devoted to observing the southern and
      northern ecliptic hemispheres, respectively, during which a vast
      discovery space was predicted \citep{sullivan2015,barclay2018}.

The survey design strategy of {\it TESS} has resulted in the
observation of stars already known to host exoplanets that were
discovered through a variety of methods. The continuous time series
photometry of these stars may be used to achieve multiple science
goals that have an over-arching theme of unprecedented
characterization of these planetary systems. These science goals
include the detection of transits for known planets
\citep{dalba2019c}, the discovery of additional planets
\citep{brakensiek2016}, the detection of phase variations and
secondary eclipses \citep{mayorga2019}, refinement of transit
ephemerides \citep{dragomir2020}, and asteroseismology of host stars
\citep{campante2016b}. Each of these science goals have been realized
to various degrees through the course of the {\it TESS} primary
mission, providing significant insight into the physical properties of
the known planets and the architectures of those systems.

In this paper, we provide a description of the science motivation
behind {\it TESS} observations of known exoplanet host stars during
the primary mission, along with statistics of these observations and a
summary of the results. Note that ``known hosts'' in this work refers
to stars that are known to host planets outside of {\it TESS}
discoveries. In Section~\ref{advantage} we present the details for
each of the science cases and quantify the advantage of returning to
known exoplanet hosts. Section~\ref{coverage} provides the statistics
of the known exoplanet host {\it TESS} observations, together with
several examples of {\it TESS} photometry for hosts observed over
multiple sectors. Section~\ref{primary} summarizes the published
science results regarding known exoplanet hosts from the {\it TESS}
primary mission, and Section~\ref{extended} discusses possible further
science yield from continuing to observe known hosts during the
extended mission. Section~\ref{conclusions} provides concluding
remarks and suggestions for additional science exploitation of known
host observations and follow-up programs.


\section{The Advantage of Observing Known Hosts}
\label{advantage}

There are numerous science motivations for observing known exoplanet
host stars. Here we discuss several of those motivations, including
transit detection of known planets, discovery of new planets, phase
signatures and secondary eclipses, transit ephemeris refinement, and
asteroseismology.


\begin{figure}
  \includegraphics[angle=270,width=8.5cm]{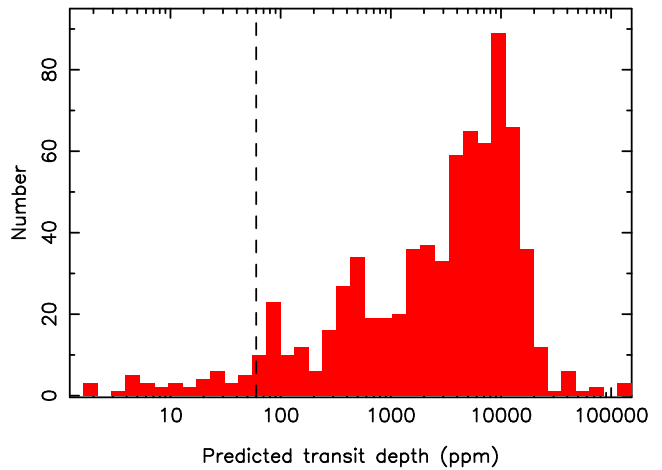}
  \caption{Histogram of predicted transit depths for all known RV
    planets. The vertical dashed line at 60~ppm represents the
    engineering requirement for the noise floor of the {\it TESS}
    photometric precision (actual noise floor is closer to 20~ppm).}
  \label{fig:transit}
\end{figure}

\subsection{Transit Detection of Known Exoplanets}
\label{transit}

At the current time, it remains unknown if many of the radial velocity
(RV) detected exoplanets transit their host stars. Since these host
stars are relatively bright, they provide numerous opportunities for
detailed characterization of the systems, such as transmission
spectroscopy, orbital dynamics, and potential targets for future
imaging missions \citep{winn2015,kane2018c,batalha2019c}. The
detection of transits for known planets has been discussed in detail
\citep{kane2007b,kane2009c,hill2020}, including the transit
probabilities of such planets \citep{kane2008b,stevens2013}. A study
of anticipated {\it TESS} observations of known exoplanet hosts was
carried out by \cite{dalba2019c}. Accounting for the transit
probability, visibility of targets, and observing cadence, this study
estimated that $11.7\pm0.3$ known RV planets would exhibit transits
during {\it TESS} primary mission observations, 3 of which would be
new transit discoveries.

Shown in Figure~\ref{fig:transit} is a histogram of predicted transit
depths for known RV planets that have not had a transit detected. The
necessary data were extracted from the NASA Exoplanet Archive
\citep{akeson2013} on 2020 May 5, and we retained all cases with the
necessary planetary and stellar information. Numerous exoplanet
mass-radius relationships have been derived
\citep{kane2012a,weiss2014,chen2017}, and we adopt the methodology of
\citet{zeng2019}, which uses a Monte Carlo approach with a planet
formation motivated growth model, to estimate planetary radii from the
minimum planetary masses. As a result, a total of 749 planets are
included in the histogram. According to \citet{ricker2015}, the
engineering requirement for the systematic noise floor of the
photometric precision over one hour timescales was $\sim$60~ppm, shown
in Figure~\ref{fig:transit} as a vertical dashed line. Of the 749
planets included, the predicted transit depths of 709 fall above this
60~ppm threshold. There are many other factors, such as additional
noise sources \citep{feinstein2019b}, geometric transit probability
\citep{kane2008b}, and transit window functions \citep{vonbraun2009},
that truncate the expected number of observed transits during {\it
  TESS} observations \citep{dalba2019c}. Fortunately, the in-flight
reassessment of the photometric precision noise floor found the
performance to be better than the engineering requirements. In most
cases, the photometric precision of {\it TESS} is sufficient to detect
transits of known RV planets should their inferior conjunction occur
during the {\it TESS} observing windows.


\subsection{Discovery of Additional Planets}
\label{additional}

One of the major reasons to continue monitoring known host stars is
the prospect of detecting additional planets within those systems,
regardless of the detection technique that was used to discover the
known planets \citep{dietrich2020}. Continued monitoring and discovery
of additional planets is an essential pathway toward revealing the
full diversity of planetary architectures \citep{winn2015}, including
dynamical interactions \citep{kane2014b,agnew2019} and coplanrity
\citep{fang2012f,becker2017c}. For example, the WASP-47 system,
initially detected as a single hot-Jupiter \citep{hellier2012}, has
been revealed as a complex multi-planet system and the focus of
numerous follow-up efforts
\citep{becker2015b,dai2015,almenara2016,sinukoff2017a,vanderburg2017,weiss2017,kane2020a}. Although
long-term photometric monitoring will only reveal those planets that
happen to have orbital alignments favorable for transit detection,
such planets typically fall within the demographic of short-period
terrestrial planets that were below the detection threshold of
previous surveys.


\subsection{Phase Variations and Secondary Eclipses}
\label{phase}

In the era of precision photometry, particularly from space-based
facilities, the detection of phase variations of exoplanets has become
a powerful method to probe atmospheric properties
\citep{faigler2011,shporer2017a}. Phase variations caused by reflected
light can provide insight into the scattering properties of an
exoplanet's atmosphere
\citep{burrows2010b,kane2010b,kane2011a,madhusudhan2012a} and can
disentangle multi-planet systems through sustained monitoring
\citep{kane2013b,gelino2014}. These reflected light signatures
complement the thermal structure and orbital information inferred from
phase variations and secondary eclipses detected in the infrared
\citep{harrington2006,knutson2007b,kane2009b,kane2011g,demory2016b}. Secondary
eclipse observations enable the measurement of atmospheric
temperatures that are critical in modeling exoplanet atmospheres and
interiors \citep{line2013c,vonparis2016a,fortney2019}. Furthermore,
the additional phase variation components of ellipsoidal variations
and Doppler beaming can be used to distinguish between stellar and
planetary companions to the host star \citep{drake2003,kane2012b}. The
{\it TESS} bandpass primarily spans optical wavelengths
\citep{ricker2015}, so the recovered phase signatures will be
dominated by the reflected light component.

\begin{figure}
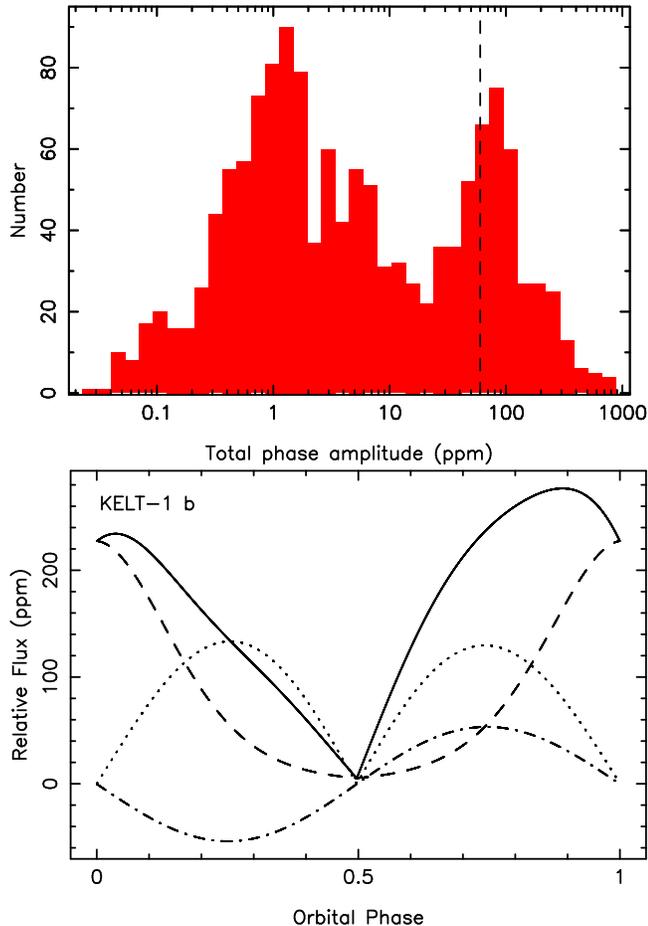

  \includegraphics[angle=270,width=8.5cm]{f02a.ps} \\
  \includegraphics[angle=270,width=8.5cm]{f02b.ps}
  \caption{Top: histogram of the total phase amplitude for all of the
    planets described in Section~\ref{phase}. The vertical dashed line
    at 60~ppm represents the noise floor of the {\it TESS} photometric
    precision. Bottom: predicted phase amplitudes for KELT-1b,
    including reflected light (dashed), ellipsoidal (dotted), Doppler
    beaming (dot-dashed), and total (solid).}
  \label{fig:phase}
\end{figure}

We used the stellar and exoplanet data from the NASA Exoplanet
Archive, as described in Section~\ref{transit}, to calculate the
reflected light, ellipsoidal variation, and Doppler beaming components
for all known planets with the necessary information. A histogram of
the combined amplitude for all three effects is shown in the top panel
of Figure~\ref{fig:phase}, including a total of 1384 planets. As for
Figure~\ref{fig:transit}, the vertical dashed line represents the
systematic noise floor of the {\it TESS} photometric precision, of
which 291 phase amplitudes lie above. For the purposes of the
reflected light calculations, the geometric albedo for all planets was
assumed to be 0.5 and we use the Keplerian orbital information where
available. The bi-model shape in the distribution arises from a
combination of exoplanet survey observational biases, and the gaps
observed in both the period and mass/size of exoplanets
\citep{matsakos2016a,mazeh2016,fulton2017}, for which the amplitudes
of the various phase components are very sensitive. In other words,
the distribution that peaks near 100~ppm is dominated by hot Jupiter
planets. For example, the predicted phase amplitudes of KELT-1b, a
$\sim$27~$M_J$ brown dwarf in a 1.22~day period orbit
\citep{siverd2012}, are represented in the bottom panel of
Figure~\ref{fig:phase}. The reflected light, ellipsoidal, and Doppler
beaming components are shown as dashed, dotted, and dot-dashed lines
respectively, and the total variations are shown as a solid
line. Orbital phase zero corresponds to a planet location of superior
conjunction, or "full" reflection phase. In this extreme case, the
combination of high mass and size, along with small star--planet
separation, results in relatiely high ampltudes for all three
components of the variations, placing it firmly within the right-hand
part of the distribution shown in the top panel of
Figure~\ref{fig:phase}.


\subsection{Transit Ephemeris Refinement}
\label{ephemeris}

\begin{figure}
  \includegraphics[width=8.5cm]{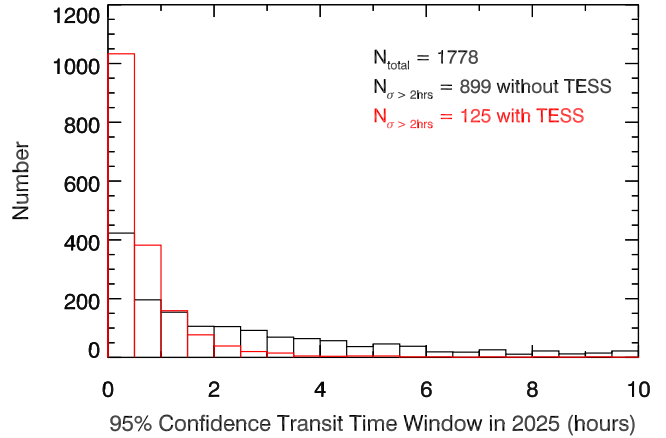}
  \caption{Histograms of the 95\% confidence window for transit times
    of 1457 known transiting exoplanets, projected forward to January
    1, 2025. The red and the black histograms are with and without
    {\it TESS} observations respectively.}
  \label{fig:ephem}
\end{figure}

The atmospheric characterization community has the ambition to study
hundreds of planets over the next decade in order to reveal the
statistics of exoplanet atmospheres. This will be largely achieved
with a combination of the James Webb Space Telescope ({\it JWST}) and
dedicated missions, such as the Atmospheric Remote-sensing Infrared
Exoplanet Large-survey ({\it ARIEL}) mission
\citep{puig2016,kempton2018}. A significant issue facing the
observational planning for atmospheric signatures of known transiting
planets is the reduced quality of their transit ephemerides with time
\citep{kane2009c}. The errors are dominated by uncertainties in the
periods, which could be significantly reduced by observing just a
handful of transits at the {\it TESS} epoch
\citep{dragomir2020,zellem2020a}. Figure~\ref{fig:ephem} shows a
histogram of the 95\% confidence window (i.e., $\pm 2\sigma$) for
transit times of 1457 well-studied transiting planets listed in the
Transiting Extrasolar Planets Catalogue \citep{southworth2011e}. The
windows were calculated for a representative date (January 1, 2025)
when {\it JWST} is expected to be in full operation. More than half of
the known planets will have windows greater than 2~hours, which means
that observations of their transits or eclipses would require
significant additional observing time to have a guaranteed
observations of a full transit event. Improvement of transit
ephemerides will be achieved via the use of various follow-up
facilities, including CHaracterizing ExOPlanets Satellite ({\it
  CHEOPS}) observations of {\it TESS} targets
\citep{broeg2014,cooke2020a}.

Some of the most exciting science from {\it Kepler} came from systems
of multiple transiting planets, particularly those where planet-planet
interactions revealed by Transit Timing Variations (TTVs) provided an
important source of mass measurements and constraints
\citep[e.g.,][]{steffen2013a,hadden2014}. Such observations of TTVs
are also true for {\it TESS} but to a more limited extent
\citep{goldberg2019,hadden2019a}. \citet{kane2019f} investigated the
degradation of TTV signals when switching from Kepler's 4-year
duration to the 6--12 month duration of {\it TESS}. Using a basic
scaling estimate, they find that roughly tens of {\it TESS} planets
will show TTVs, although only some of these will lead to useful mass
constraints.


\subsection{Asteroseismology}
\label{asteroseismology}

\begin{figure}
  \includegraphics[width=8.5cm,clip]{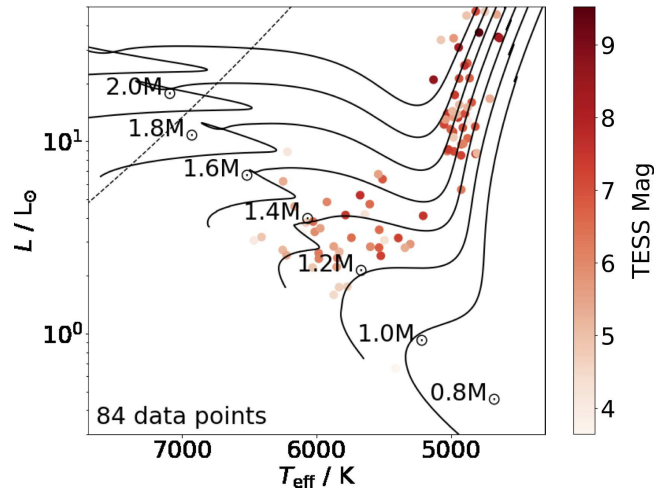}
  \caption{Predicted yield of known host stars in Cycles 1 and 2
    having detectable solar-like oscillations. Solar-calibrated
    evolutionary tracks span the mass range 0.8--2.0~$M_\odot$. The
    slanting dashed line represents the red edge of the $\delta$~Scuti
    instability strip. Evolved stars (i.e., with $\log g < 3.85$) make
    up about half of the yield.}
  \label{fig:HRseismo}
\end{figure}

Asteroseismology is one of the most successful methods to precisely
infer radii, masses, and ages of exoplanets through the
characterization of their host stars \citep[for recent reviews,
  see][]{huber2018,lundkvist2018}. We predicted the asteroseismic
yield of known host stars in Cycles 1 and 2 by employing a statistical
test \citep{chaplin2011b,campante2016b,schofield2019} that estimates
the detectability of convection-driven, solar-like oscillations in
{\it TESS} photometry of any given target. The expectation is that
solar-like oscillations are detectable in nearly 100 solar-type (i.e.,
low-mass, main-sequence stars and cool subgiants) and red-giant known
hosts, virtually all of which are RV systems (see Figure
\ref{fig:HRseismo}). Moreover, about half of such hosts are evolved
stars, i.e., having $\log g < 3.85$. The corresponding planet sample
is mostly comprised of long-period gas giants, with a smaller fraction
of hot Jupiters and warm super-Earths/Neptunes. To assess if
asteroseismology can further constrain stellar and planetary
properties, we estimated the precision with which fundamental stellar
properties can be obtained for stars in the asteroseismic sample. We
used the Bayesian code PARAM
\citep{dasilva2006b,rodrigues2014,rodrigues2017} to this end, a
grid-based approach whereby observables are matched to well-sampled
grids of stellar evolutionary models. Two different sets of
observables were considered, one containing only spectroscopic data
and a parallax-based luminosity (this allows reproducing the typical
precision levels currently found in the literature), the other
containing additional constraints from asteroseismology (namely, the
predicted large frequency separation, $\Delta\nu$, and the predicted
frequency of maximum oscillation amplitude, $\nu_{\rm max}$, both with
uncertainties as expected for {\it TESS}). We found that by including
asteroseismic constraints one can significantly improve (by a factor
of 2--5) the precision of stellar properties when compared to
estimates stemming from a combination of spectroscopy and astrometry
alone (1.9 vs 3.4\% in radius, 4.6 vs 6.7\% in mass, 15 vs 30\% in
age, and 3.4 vs 15\% in mean density). This asteroseismic sample will
thus provide us with a benchmark ensemble of planets with precisely
inferred radii, masses, and ages.

\section{Known Host Coverage}
\label{coverage}

The nominal plan for {\it TESS} observations during the primary
mission was to result in $\sim$85\% sky coverage with a minimum
observing baseline of $\sim$27~days \citep{ricker2015}. Year 1 (Cycle
1) and year 2 (Cycle 2) of the mission were directed at the southern
and northern ecliptic hemispheres, respectively. During Cycle 1,
modifications were made to the location of the Cycle 2 sectors that
shifted them north along a line of ecliptic longitude in order to
minimize scattered light effects\footnote{\tt
  https://tess.mit.edu/observations/}.

For our analysis of the {\it TESS} coverage of known exoplanet hosts
during the primary mission, we include data from the NASA Exoplanet
Archive \citep{akeson2013} from 2020 May 5, matching the sample
described in Section~\ref{transit} and Section~\ref{phase}. From these
data, we exclude those exoplanet hosts whose planets were detected by
{\it TESS} (45) and host stars without $V$ magnitude information
(202). These restrictions reduce the planet sample from 4152 to
3912. For each host star, we determined the sectors during which they
were observed by {\it TESS} using the proposal tools provided by the
{\it TESS} Science Support Center\footnote{\tt
  https://heasarc.gsfc.nasa.gov/docs/tess/}.

\begin{figure*}
  \begin{center}
    \begin{tabular}{cc}
      \includegraphics[width=8.5cm]{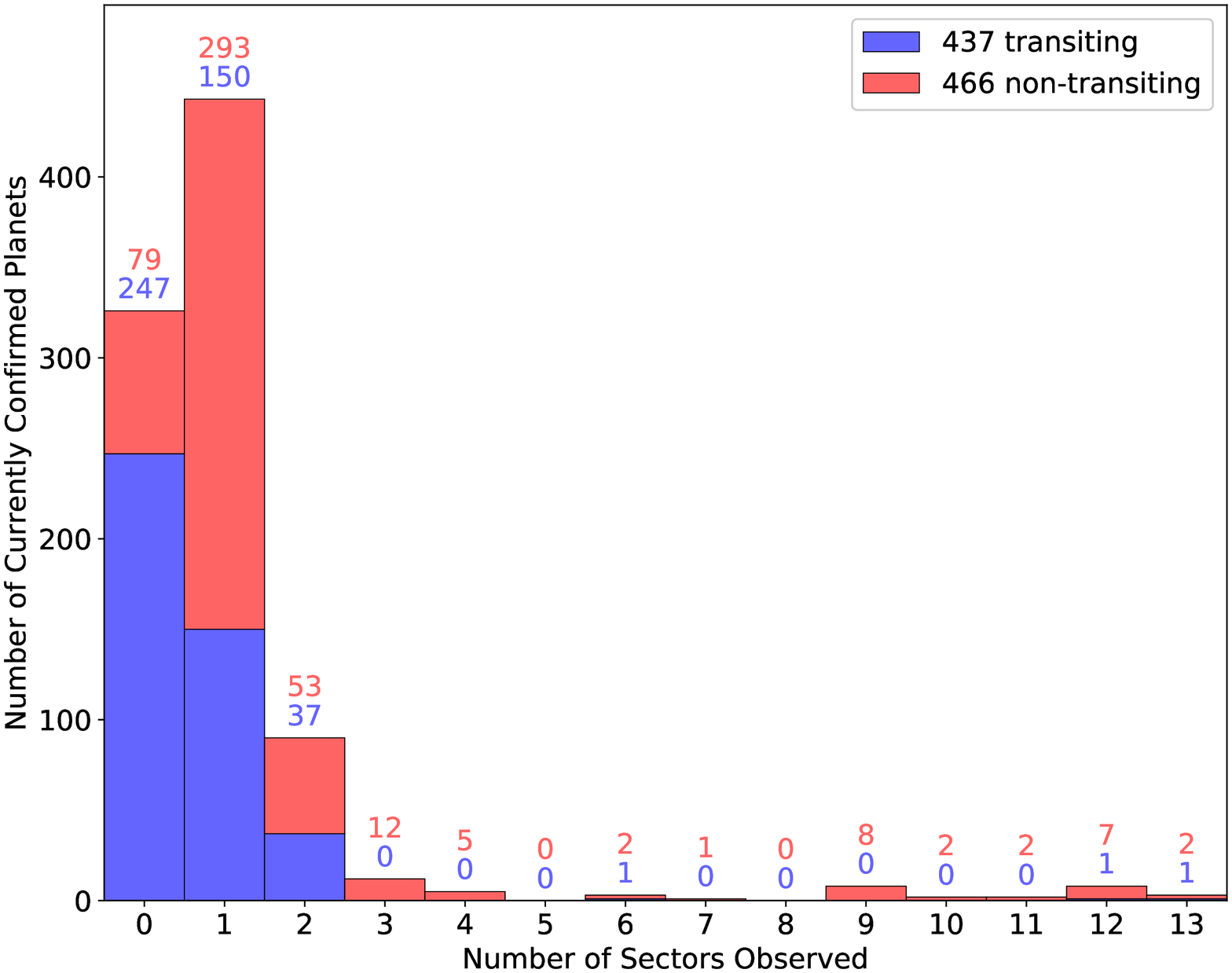} &
      \includegraphics[width=8.5cm]{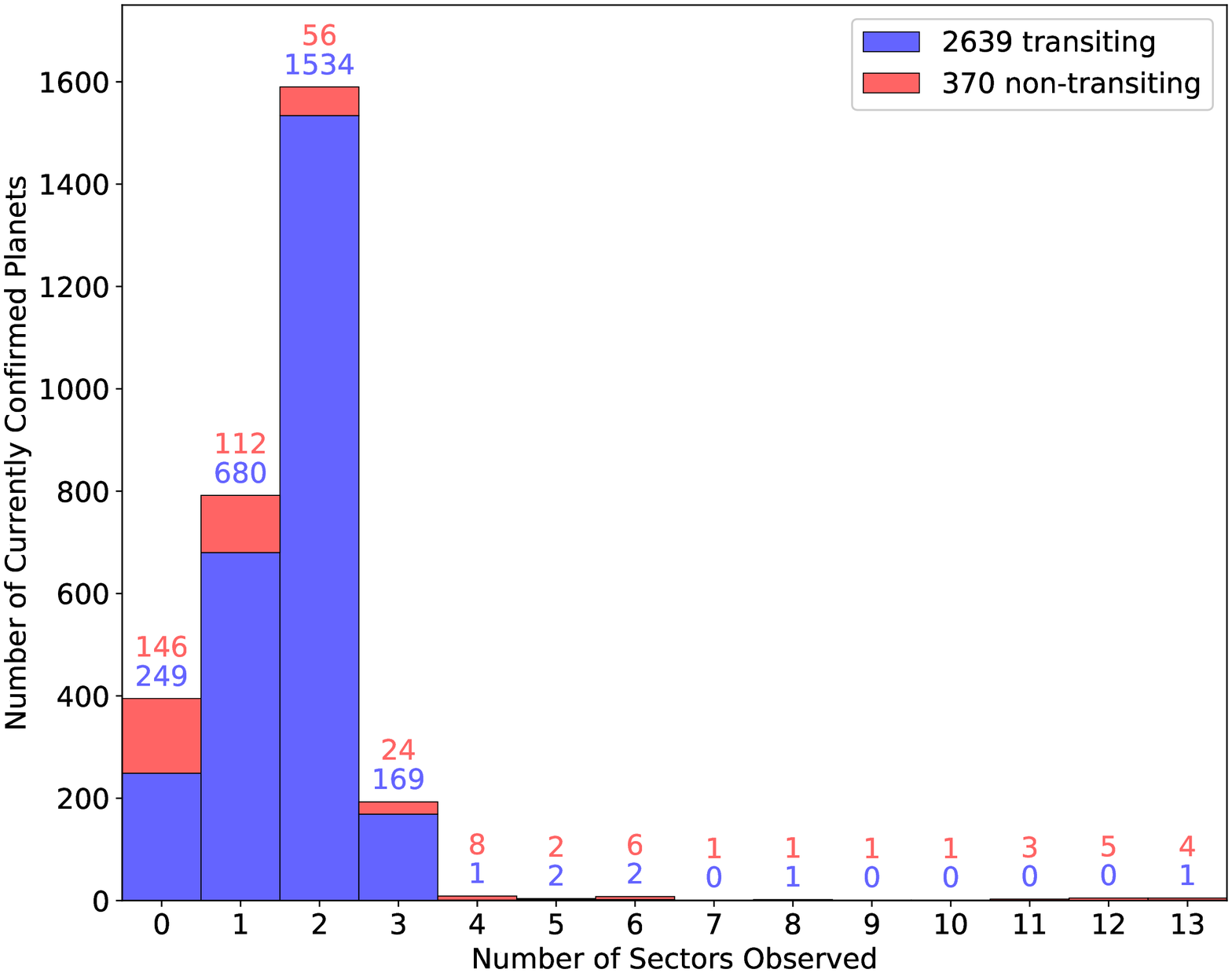} \\
      \includegraphics[width=8.5cm]{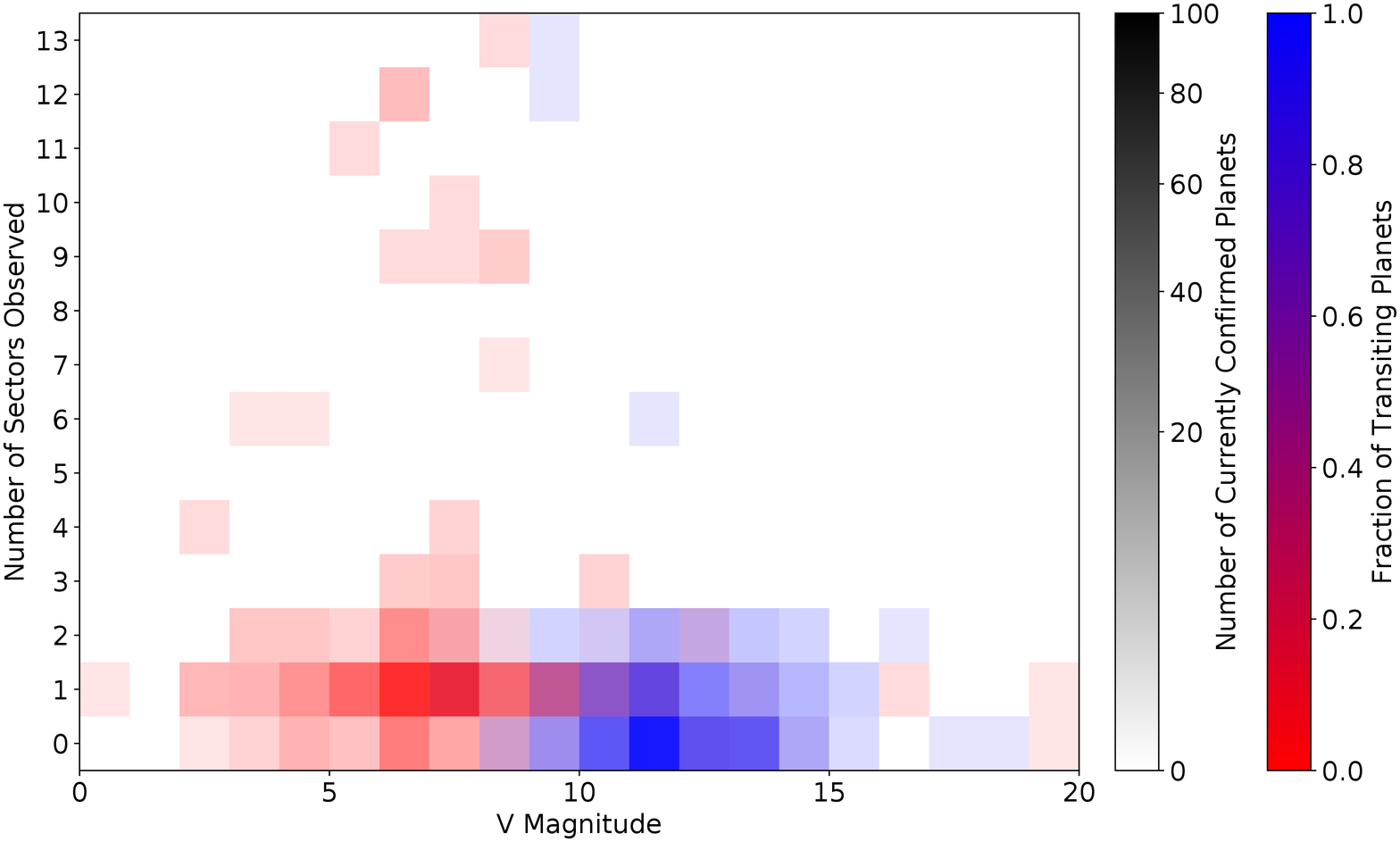} &
      \includegraphics[width=8.5cm]{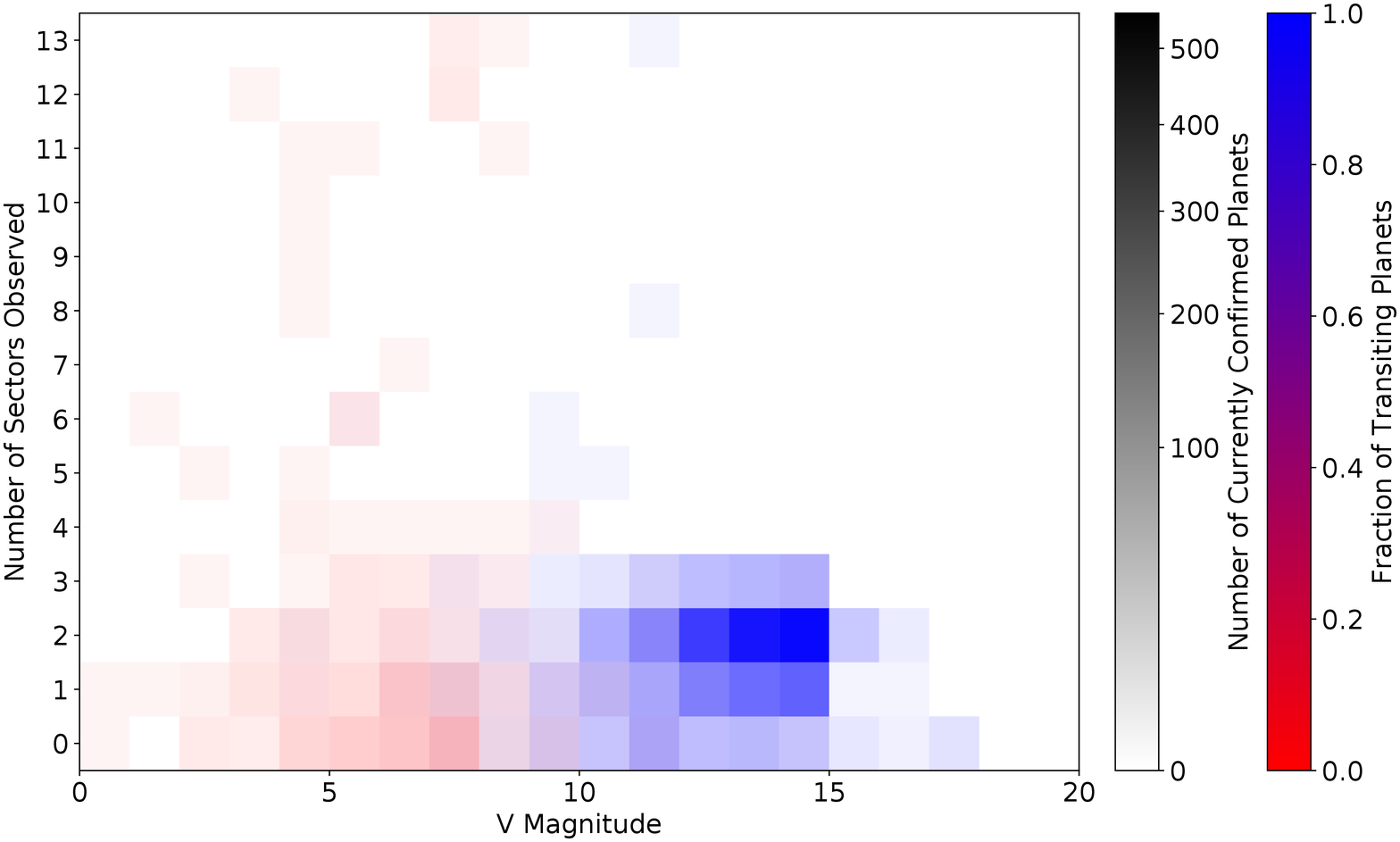} \\
    \end{tabular}
  \end{center}
  \caption{{\it TESS} coverage of the known exoplanets during the
    primary mission. The top-left and top-right panels show histograms
    of the number of sectors during which transiting (blue) and
    non-transiting (red) planets were covered during Cycle 1 and Cycle
    2, respectively. The bottom-left and bottom-right panels show
    intensity maps of the known exoplanet coverage as a function of
    the host star $V$ magnitude during Cycle 1 and Cycle 2,
    respectively.}
  \label{fig:coverage}
\end{figure*}

Shown in Figure~\ref{fig:coverage} are plots that represent the {\it
  TESS} coverage of the known exoplanet hosts during the primary
mission. The two left-hand panels in Figure~\ref{fig:coverage} refer
to Cycle 1 and the two right-hand panels refer to Cycle 2. The top two
panels are histograms of the total number of sectors during that cycle
for which exoplanets were covered by {\it TESS} observations, with
transiting planets being shown in blue and planets not known to
transit shown in red. The numbers above each bin indicate the number
of planets represented by the transiting and non-transiting categories
for that bin. Although the number of exoplanets covered during Cycle 1
are fairly evenly split between the transiting and non-transiting
categories, the exoplanet sample in Cycle 2 was dominated by the
observations of the {\it Kepler} field, most of which are too faint
for {\it TESS} data to be profitable. We estimate the fraction of
known exoplanet hosts covered by the {\it TESS} primary mission by
removing those exoplanets with zero sector coverage (see bin 0 of the
Figure~\ref{fig:coverage} histograms) from the total number of
exoplanets in our sample (3912). This results in a fractional
exoplanet host coverage of $\sim$81.5\%.

The bottom two panels of Figure~\ref{fig:coverage} present the same
data as for the top two panels, but in the form of intensity maps as a
function of both sectors observed and the $V$ magnitude of the host
stars. The shading and color of the bins relate to the number of
planets in that bin and the relative fractions of transiting
planets. These bottom two plots of Figure~\ref{fig:coverage} emphasize
the bimodal distribution of host star $V$ magnitudes between RV and
transit surveys, resulting from the need of transit surveys for large
stellar samples to overcome the geometric transit probability, thus
including many more fainter stars than brighter stars
\citep{kane2009c}. As for the top-right panel, the {\it Kepler} sample
dominates the data shown in the bottom-right panel, causing an
apparent lack of contrast in the intensity map.

Figure~\ref{fig:coverage} indicates that a handful of known hosts were
observed almost continuously during a given cycle of {\it TESS}
observations. For example, consider the HD~40307 system, which was
observed for 12 of the 13 sectors of Cycle 1. The system is known to
contain at least 5 planets that are a mixture of super-Earths and
mini-Neptunes discovered using the RV technique
\citep{mayor2009a,tuomi2013a,diaz2016a}. Currently, none of the
planets are known to transit the host star, a K2.5 dwarf
\citep{tuomi2013a}. To calculate the transit probabilities and
predicted transit depths, we combined the minimum planet masses of
\citet{tuomi2013a} with the mass-radius relationship of
\citet{chen2017} to estimate radii of 1.8, 2.5, 3.0, 2.1, and
2.6~$R_\oplus$ for the b, c, d, f, and g planets, respectively. We
further adopted the stellar radius estimate of $R_\star =
0.7083$~$R_\odot$ provided by \citet{valenti2005}. These result in
transit probabilities of 11.2\%, 6.5\%, 4.0\%, 2.1\%, and 0.9\% for
the b, c, d, f, and g planets, respectively. Note that these
probabilities are calculated independently of each other and do not
take into account coplanarity of the system. The calculated predicted
transit depths are 215, 415, 598, 293, and 449 ppm for the b, c, d, f,
and g planets, respectively.

\begin{figure*}
  \begin{center}
    \includegraphics[width=18.0cm]{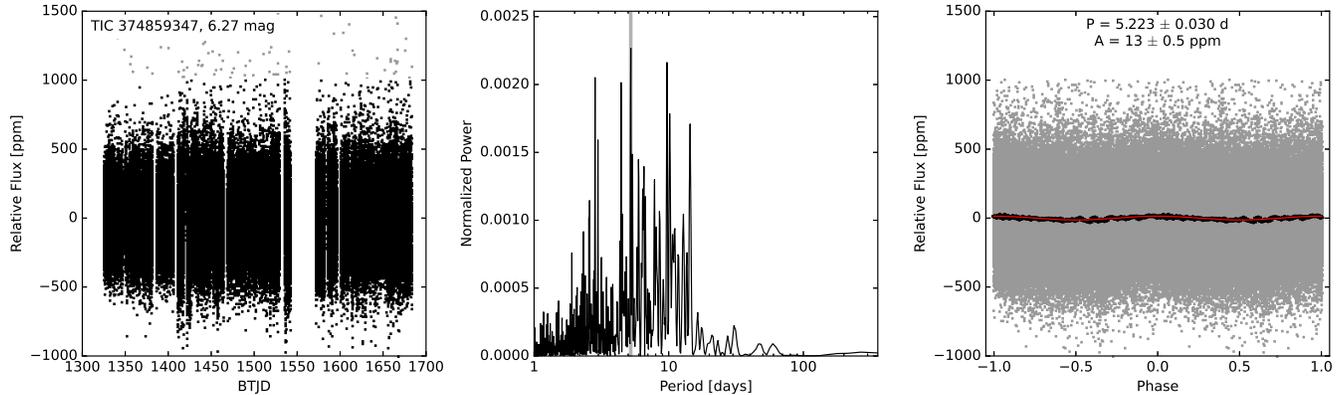}
  \end{center}
  \caption{{\it TESS} observations of the known exoplanet host
    HD~40307. Left: light curve from 12 sectors of {\it TESS}
    observations during Cycle 1. Middle: a Lomb-Scargle periodogram
    calculated from the {\it TESS} photometry shown in the left
    panel. Right: photometry folded on the most significant period
    detected from the photometric variability analysis (5.223~days).}
  \label{fig:hd40307}
\end{figure*}

Shown in Figure~\ref{fig:hd40307} are the {\it TESS} photometry for
HD~40307, with a 1$\sigma$ scatter of 203~ppm, and the results of a
variability analysis of the data. The dates shown in the left panel
are expressed in Barycentric {\it TESS} Julian Day (BTJD), where BTJD
= BJD $-$ 2457000. We used the Presearch Data Conditioning (PDC)
photometry, processed by the Science Processing Operations Center
(SPOC) pipeline
\citep{smith2012d,stumpe2012,stumpe2014,jenkins2016,jenkins2020a}, and
we extracted the data using the {\sc Lightkurve} tool
\citep{lightkurve2018}. The precision of the data is sufficient to
rule out the previously calculated transits depths for all 5
planets. Even though the star was not observed during Sector 9, the
longest period planet ($\sim$197~days) is sufficiently covered during
Sectors 1--8 that the predicted 449~ppm for that planet can also be
excluded from the data. An alternative explanation is that the planets
do transit but their bulk densities are significantly higher than that
predicted from typical mass-radius relationships. Examples such as the
case of HD~40307 demonstrate the power of {\it TESS} to systematically
achieve dispositive null detections of transits that are exceptionally
difficult to achieve from ground-based observations \citep{wang2012}.


\section{Science From the Primary Mission}
\label{primary}

Observations of known exoplanet hosts during the {\it TESS} primary
mission have realized many of the goals described in
Section~\ref{advantage}. Here we outline the major discoveries that
have occurred in each of the Section~\ref{advantage} categories.

{\it Transits of known planets (Section~\ref{transit}).} A total of
three known RV planets were discovered to transit from {\it TESS}
observations during the primary mission. These include HD~118203b, a
Jovian planet in a 6.13~day orbit \citep{pepper2020}, and HD~136352 b
and c, a super-Earth and mini-Neptune in 11.6~day and 27.6~day orbits,
respectively \citep{kane2020c}. The number of RV planets found to
transit during the primary mission is aligned with the predictions of
\citet{dalba2019c}, which predicted three such discoveries.

{\it New planets in known systems (Section~\ref{additional}).} An
early science result from {\it TESS} observations was the detection of
an additional inner transiting planet in the Pi Mensae system
\citep{huang2018}. The combination of a Jovian planet in an eccentric
5.7~year period orbit with a mini-Neptune in a 6.27~day period orbit
makes the system of dynamical interest
\citep{derosa2020c,xuan2020b}. Similarly, the long-period
($\sim$1600~days) Jovian planet in the HD~86226 system was found by
\citet{teske2020} to host a transiting mini-Neptune planet in a
3.98~day orbit.

{\it Phase variations (Section~\ref{phase}).} Numerous known
transiting planets have been the subject of phase variation studies to
place important constraints on their atmospheric properties. These
include WASP-18b \citep{shporer2019}, WASP-19b \citep{wong2020b},
WASP-121b \citep{daylan2020a}, KELT-1b \citep{beatty2020}, and KELT-9b
\citep{wong2020d}. Note that KELT-9b also exhibited an asymmetric
transit in {\it TESS} photometry that was caused by rapid stellar
rotation combined with a spin-orbit misalignment
\citep{ahlers2020b}. A systematic study of phase curves detected for
known transiting planets during Cycle 1 was carried out by
\citet{wong2020e}.

{\it Transit ephemeris refinement (Section~\ref{ephemeris}).} As
described earlier, the refinement of transit ephemerides is a crucial
component for enabling valuable follow-up observations, particularly
those that involve atmospheric characterization \citep{kempton2018}. A
concerted effort has been undertaken by various teams to combine {\it
  TESS} data with ground-based observations
\citep{yao2019,corteszuleta2020,edwards2020a} and {\it K2} data
\citep{ikwutukwa2020} to improve the orbital properties of known
transiting planets. Additionally, unexpected variations in the transit
times of WASP-4b were detected by \citet{bouma2019a} and confirmed by
\citet{southworth2019}, and were later explained by acceleration
effects of the WASP-4 system \citep{bouma2020a}.

{\it Stellar characterization through asteroseismology
  (Section~\ref{asteroseismology}).} Several known exoplanet hosts
have benefited from the {\it TESS} precision photometry during the
primary mission, particularly those that have evolved past the main
sequence. \citet{campante2019} reported the detection of solar-like
oscillations in the light curves of the red-giant exoplanet hosts
HD~212771 and HD~203949. A further detection of solar-like
oscillations was reported by \citet{jiang2020} for the giant host star
HD~222076, greatly improving the determined mass, radius, and age of
the star. \citet{nielsen2020d} used {\it TESS} asteroseismology to
firmly place the well-studied host $\lambda^2$ Fornacis at the early
stage of its subgiant evolutionary phase.


\section{Extended Mission Science Yield}
\label{extended}

\begin{figure}
  \includegraphics[width=8.5cm]{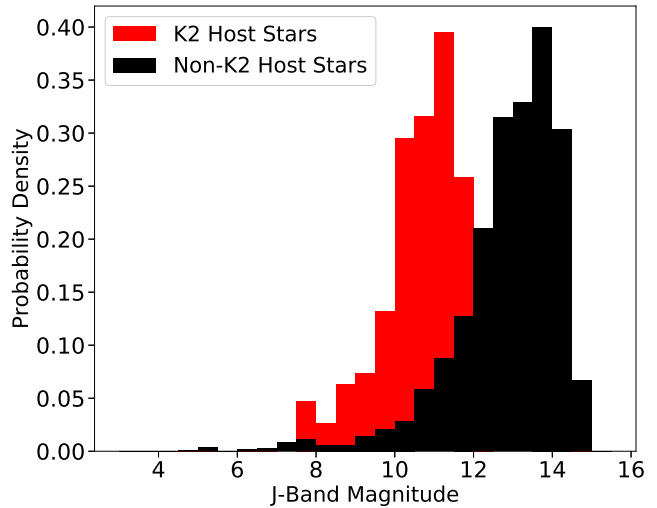}
  \caption{Histograms of host star $J$ magnitudes for exoplanets
    discovered via the transit method. Shown in red are those
    discovered with {\it K2}, and shown in black are those from all
    other transit surveys.}
  \label{fig:k2}
\end{figure}

{\it TESS} has now moved in to the extended mission, from which
further observations of known exoplanet host stars will result. For
transits of known RV exoplanets, \citet{dalba2019c} predict that {\it
  TESS} will reveal one such planet be transiting for each year of the
extended mission during which it returns to one of the hemispheres
observed during the primary mission. As described in
Section~\ref{transit}, the RV host stars are generally brighter than
those of transit surveys, and so are valuable targets for follow-up
observations. Likewise, extending the observations baseline for known
systems, both transiting and non-transiting, will undoubtedly reveal
further planets in those systems, adding to our statistical knowledge
of planetary architectures. For the phase variations science, the
advantage of returning to previously observed fields is to build
signal-to-noise for small planet phase signatures that may have had an
initial tenuous detection. The probability of detecting solar-like
oscillations for a given star depends sensitively on the length of the
observations \citep{chaplin2011b,campante2016b,schofield2019}. As the
baseline increases, so will the relative statistical fluctuations in
the underlying background power in the Fourier spectrum decrease in
magnitude. Consequently, further {\it TESS} observations of known
hosts (even when the data are not contiguous) will allow the
confirmation of previous tentative detections of oscillations as well
as providing new detections.

Cycle 3 for {\it TESS} observations are returning to the southern
ecliptic hemisphere, complementing the prior observations of the same
stars during Cycle 1. Beyond Cycle 3, it is expected that {\it TESS}
observations will turn to the ecliptic, observing stars not previously
measured during the mission.  Furthermore, the ecliptic observations
will be carried out with the spacecraft rotated by 90$\degr$ relative
to the nominal pointing configuration. Such an observing strategy will
cause a partial overlap of the camera fields with previously observed
sectors in the northern and southern ecliptic, and significant overlap
of with other ecliptic fields. This overlap of the ecliptic fields
will result in a longer time baseline of observations relative to the
$\sim$27~day duration for most of the camera pointings during the
primary mission, allowing a much greater sensitivity to longer period
planets and a higher science yield for many of the known host science
cases described in this work. Another factor in favor of ecliptic
observations are the relative brightness of the {\it K2} mission host
stars (whose transit discovery fields were largely centered along the
ecliptic) and the subsequent potential for science
return. Figure~\ref{fig:k2} shows histograms of the host star $J$
magnitude for planets that were discovered using the transit
method. The histograms are for those cases discovered by {\it K2} and
those discovered via all other transit surveys (once again, excluding
{\it TESS}). There is a clear bi-modality in the overall host star
brightness distribution in which the {\it K2} host stars are
preferentially brighter. One effect of this brightness distribution is
that {\it K2} discoveries are more likely to result in successful
atmospheric characterization studies \citep{kosiarek2019a}. To
demonstrate this proposition, we used the Transmission Spectroscopy
Metric (TSM) devised by \citet{kempton2018}, and recently applied to
{\it TESS} exoplanet candidates \citep{ostberg2019}. We calculated the
TSM for all exoplanets with available data for the {\it K2} and
non-{\it K2} groups represented in Figure~\ref{fig:k2}. Note that
faint host stars are less likely to have mass measurements for their
planets (a required component of the TSM calculation), so in those
cases we estimated the planet mass using the methodology of
\citet{chen2017}. These TSM calculations revealed a mean value of 25.2
for the {\it K2} population and 8.7 for the non-{\it K2}
population. Thus, {\it TESS} observations of the {\it K2} host stars
along the ecliptic would provide enormous benefits for the transit
ephmeris refinement described in Section~\ref{ephemeris} in
preparation for potential follow-up observations.


\section{Conclusions}
\label{conclusions}

The {\it TESS} mission has completed a highly successful survey of the
sky during the first two years. Although the discovery of previously
unknown planetary systems is the primary science goal of the mission,
{\it TESS} has provided serendipitous insights into previously known
systems, aiding toward the characterization of some of the brightest
and well-known host stars. As we have demonstrated here, $\sim$81.5\%
of known exoplanet hosts were observed during the primary mission, of
which most of those outside the {\it Kepler} field were observed for a
single sector. Regardless, the science yield for these targets was
extensive, covering a broad range of topics.

The significant discoveries include transit detection of known RV
planets orbiting nearby and bright host stars, such as the naked-eye
star HD~136352, and additional transiting planets in known
exosystems. The combination of precise photometry with the relatively
bright exoplanet hosts of known transiting planets has enabled
substantial progress to be made in the detection of phase variations,
providing further constraints on the atmospheric properties for these
planets. Observations of these known transiting systems has also
greatly improved the precision of their measured orbital parameters,
which is a critical factor in scheduled follow-up observations with
large competitive facilities. Finally, the observation of evolved
hosts by {\it TESS} has made it possible to greatly improve the
properties of these stars, including mass, radius, and age, and thus
better understand the planets that orbit them.

Beyond the primary mission, it is expected that further {\it TESS}
observations of known exoplanet hosts will continue to yield exciting
new results as the baseline of observations is increased and new
fields along the ecliptic are covered. In particular, the extended
baseline will likely reveal further transits of known and unknown
planets for stars already known to harbor planets. As we have shown,
these systems have preferentially bright host stars and will form a
major contribution to the target selection for atmospheric
characterization observations. Thus, an important component of the
{\it TESS} legacy will be to establish the cornerstone systems whose
history and future of observations place them amongst our best
understood examples of planetary systems outside of the solar system.


\section*{Acknowledgements}

The authors would like to thank Darin Ragozzine for his contributions
to the Guest Investigator program. S.R.K. acknowledges support by the
National Aeronautics and Space Administration through the TESS Guest
Investigator Program (17-TESS17C-1-0004). P.D. acknowledges support
from a National Science Foundation Astronomy and Astrophysics
Postdoctoral Fellowship under award AST-1903811. T.L.C. acknowledges
support from the European Union's Horizon 2020 research and innovation
programme under the Marie Sk\l{}odowska-Curie grant agreement
No.~792848 (PULSATION). Funding for the {\it TESS} mission is provided
by NASA's Science Mission directorate. This research has made use of
the Exoplanet Follow-up Observation Program website, which is operated
by the California Institute of Technology, under contract with the
National Aeronautics and Space Administration under the Exoplanet
Exploration Program. Resources supporting this work were provided by
the NASA High-End Computing (HEC) Program through the NASA Advanced
Supercomputing (NAS) Division at Ames Research Center for the
production of the SPOC data products. This paper includes data
collected by the TESS mission, which are publicly available from the
Mikulski Archive for Space Telescopes (MAST). This research made use
of Lightkurve, a Python package for Kepler and TESS data
analysis. This research has made use of the NASA Exoplanet Archive,
which is operated by the California Institute of Technology, under
contract with the National Aeronautics and Space Administration under
the Exoplanet Exploration Program. The results reported herein
benefited from collaborations and/or information exchange within
NASA's Nexus for Exoplanet System Science (NExSS) research
coordination network sponsored by NASA's Science Mission Directorate.


\software{Lightkurve \citep{lightkurve2018}}




\end{document}